\documentclass[aps, 10pt, superscriptaddress, twocolumn,groupedaddress]{revtex4-1}
	\usepackage[bookmarks=false,linkcolor=blue,urlcolor=blue,colorlinks,citecolor=blue]{hyperref}

	\usepackage{graphicx}
	\usepackage{color}
	\usepackage{amsmath}
	\usepackage[normalem]{ulem}
	\usepackage[toc,page]{appendix}
\usepackage{amsmath,dsfont,url,amssymb}
\usepackage{overpic}

\newcommand{\be}{\begin{equation}}
\newcommand{\ee}{\end{equation}}
\newcommand\p{\ensuremath{\partial}}
\newcommand\vep{\varepsilon}

\definecolor{inkred}{RGB}{210,29,0}
\definecolor{inkblue}{RGB}{0,112,196}

\def \d {\partial}

\DeclareMathOperator{\Tr}{Tr}

\renewcommand{\Im}[0]{\operatorname{Im}}
\renewcommand{\Re}[0]{\operatorname{Re}}

\begin{document}

\title{\begin{flushright}\vspace{-1in}
			\mbox{\normalsize  EFI-20-4}
		\end{flushright}
	Breakdown of Diffusion on Chiral Edges \vspace{2pt}
	 }

\author{Luca V.~Delacr\'etaz}
\author{Paolo Glorioso \vspace{5pt}}

\affiliation{Kadanoff Center for Theoretical Physics, University of Chicago, Chicago, IL 60637 \vspace{4pt}}

\preprint{EFI20-4}

\begin{abstract}
We show that dirty Quantum Hall systems exhibit large hydrodynamic fluctuations at their edge that lead to anomalously damped charge excitations in the Kardar-Parisi-Zhang universality class $\omega \simeq ck - i \mathcal D k^{3/2} $. The dissipative optical conductivity of the edge is singular at low frequencies $\sigma(\omega) \sim 1/\omega^{1/3}$. These results are direct consequences of the charge continuity relation, the chiral anomaly, and thermalization  on the edge -- in particular translation invariance is not assumed. Diffusion of heat similarly breaks down, with a universality class that depends on whether the bulk thermal Hall conductivity vanishes. We further establish the theory of fluctuating hydrodynamics for surface chiral metals, where charge fluctuations give logarithmic corrections to transport.
\end{abstract}


\maketitle


\section{Introduction and Results}

Quantum Hall (QH)  droplets feature gapless excitations at their edge  \cite{PhysRevB.41.12838}. At temperatures far below the bulk gap, the bulk essentially remains non-dissipative but the edge is expected to thermalize; thermalization implies that modes not protected by conservation laws should relax. In particular, the plethora of chiral Luttinger liquid  channels predicted for certain QH states are damped by disorder and interactions, and only the collective excitations corresponding to charge \cite{PhysRevLett.72.4129,PhysRevB.51.13449} and heat \cite{PhysRevB.55.15832} survive at late times. Early experiments in GaAs \cite{PhysRevB.45.3894,PhysRevB.50.1582} indeed observed a single linearly dispersing collective excitation -- the edge magnetoplasmon, associated with charge fluctuations -- and later experiments found evidence for the neutral heat mode \cite{Bid2010,Venkatachalam2012}. More recently, these QH edge modes were observed in graphene \cite{PhysRevLett.110.016801,PhysRevLett.113.266601} and cold atoms \cite{Goldman6736,Stuhl1514}.

Charge propagates ballistically on the edge, in the direction fixed by the sign of the filling $\nu=n/B$.
The damping of this mode was first studied at zero temperature in Ref.~\cite{volkov1988edge}. In the hydrodynamic regime, i.e. at finite temperature $T$ and low frequencies $\omega\tau_{\rm th} \ll 1$, the chiral ballistic front is expected to broaden diffusively \cite{PhysRevB.51.13449}. The thermalization time $\tau_{\rm th}$ may be controlled by various mechanisms depending on the microscopics of the edge %
	\footnote{One possible mechanism for thermalization at the edge of fractional QH bulks is interchannel scattering in the presence of disorder \cite{PhysRevLett.72.4129,PhysRevB.51.13449} (see e.g.~Refs.~\cite{PhysRevB.85.041102,PhysRevB.92.085124} for other possible mechanisms, still in the context of Luttinger liquids). However even at integer fillings edge electrons feel Coulomb interactions and may thermalize. }
-- the central assumption in this Letter is that it is sufficiently small so that frequencies $\omega \lesssim 1/\tau_{\rm th}$ can be probed experimentally.

Using fluctuating hydrodynamics, we will find that nonlinearities are relevant; large charge fluctuations lead to a breakdown of diffusion and drive the edge to a dissipative fixed point in the Burgers-Kardar-Parisi-Zhang (KPZ) universality class \cite{PhysRevA.16.732,PhysRevLett.56.889}, with dynamic critical exponent $z = 3/2$ controlling the broadening of the chiral ballistic front $\omega \simeq c k - i \mathcal D k^{3/2}$.
Breakdown of diffusion leads to a failure of the Einstein relation, and the optical conductivity is singular at low frequency $\sigma(\omega)\sim 1/\omega^{1/3}$.
Singular low frequency transport is a hallmark of large hydrodynamic fluctuations \cite{PhysRevA.16.732,PhysRevLett.89.200601}: when hydrodynamic interactions are instead irrelevant, response functions at the lowest frequencies are analytic and the interesting physics is instead hidden e.g.~in the temperature dependence of transport parameters. We stress that momentum conservation is not assumed -- disorder therefore does not have to be introduced by hand, and does not regulate the singularity in $\sigma(\omega)$ which is only cut off by finite system size.

Without momentum conservation, hydrodynamic fluctuations are usually irrelevant and give small `long-time tail' corrections to diffusive transport \cite{Ernst1984, PhysRevB.73.035113} $\sigma(\omega) = \chi D + |\omega|^{d/2}$, where $D$ is the diffusion constant and $d$ the spatial dimension. The difference here stems from the fact that the $U(1)$ symmetry has a chiral anomaly. The interplay of anomalies and hydrodynamics has been appreciated since the work of Son and Surowka \cite{Son:2009tf}. Although anomalies often only lead to subtle effects on transport, we show that the (1+1)d chiral anomaly has dramatic consequences, with ballistic propagation and large hydrodynamic fluctuations.

The connection between momentum-conserving hydrodynamics and the KPZ universality class has been long known \cite{PhysRevA.16.732,PhysRevLett.89.200601,Spohn2014}. More recently it was shown that the hydrodynamics of a non-integrable spin chain, despite the lack momentum conservation, shows KPZ scaling at intermediate energies \cite{Das2019}. We show here that systems with a chiral anomaly display KPZ scaling all the way down to arbitrarily low energies even without momentum conservation.

{
Our main results quoted above for charge fluctuations are obtained in Section \ref{sec_main}; energy fluctuations are then studied in Section \ref{sec_heat}, and diffusion of heat on the edge is similarly shown to break down. In Section \ref{sec_surface}, we show that the KPZ fixed point describing the chiral edge can be accessed perturbatively from the upper critical dimension $d_c=2$, where interactions are marginally irrelevant. Incidentally, the theory at $d=d_c$ is interesting in its own right because it describes the hydrodynamics of surface chiral metals \cite{PhysRevLett.76.2782,PhysRevLett.75.4496}, i.e.~coupled layered edge states. This generalization to higher dimensions is distinct from the one natural for the KPZ equation, where interactions are instead marginally relevant in $d=2$.

}

\section{Charge fluctuations on the edge}\label{sec_main}

We study systems in one spatial dimension with a single $U(1)$ symmetry, that is anomalous
\begin{equation}\label{eq_anom}
\d_\mu j^\mu = \frac{\nu}{4\pi} \epsilon^{\mu\rho}F_{\mu\rho}\, .
\end{equation}
We are working in units where $e^2/\hbar = 1$. Such systems can be thought of as living on the boundary of a gapped bulk. When $\nu\in \mathbb Z$, the topological order in the bulk is trivial and the anomaly can be canceled by a properly quantized Chern-Simons term $\frac{\nu}{4\pi} A d A$ for the background field. When $\nu \notin \mathbb Z$ as in fractional QH states, the bulk has non-trivial topological order. We make no additional symmetry assumptions -- in particular, momentum is not approximately conserved in any limit.

We are interested in the finite temperature properties of the system on the edge, at temperatures small compared to the bulk gap $k_BT\ll \Delta$. We will assume that the edge thermalizes -- this implies that physics at the lowest frequencies is governed by hydrodynamics, namely the dynamics of conserved densities: charge $n=j^0$, and heat (or energy). We postpone the treatment of heat to the next section; as we will see the dynamics of charge density alone is already surprisingly rich.

Dissipation in a theory with a non-anomalous $U(1)$ symmetry is described by simple diffusion $\omega\sim - i  D k^2$. The goal of this section is to determine how the anomaly $\nu\neq 0$ changes this picture. The hydrodynamic treatment proceeds as follows  \cite{KADANOFF1963419,forster2018hydrodynamic}: at late times, all operators are controlled by their overlaps with hydrodynamic densities, here $n$. This holds in particular for the current, which can be written in terms of $n$ -- or its associated potential $\mu$ -- in a gradient expansion
\begin{equation}\label{eq_consti}
j_x
	= \frac{\nu}{2\pi} \hspace{1pt} \mu - \chi D \d_x \mu + \cdots\, ,
\end{equation}
where the charge susceptibility $\chi=\d n / \d\mu$ and diffusivity $D$ are unknown functions of $n$ (or $\mu$), and $\cdots$ denotes higher gradient terms $O(\d_x^2 \mu)$. The anomaly fixes the leading term in the constitutive relation %
	\footnote{This is analogous to how the chiral anomaly in (3+1)d hydrodynamics fixes terms in the constitutive relation that are first order in gradients \cite{Son:2009tf}. The anomaly has a more important role here, as it enters at zeroth order in the expansion. A similar zeroth order anomaly fixes the speed of sound of superfluids \cite{Delacretaz:2019brr}.}.
Combining Eqs.~\eqref{eq_anom} and \eqref{eq_consti}, one finds the following equation of motion for the charge density
\begin{equation}\label{eq_eom}
0 = \dot n + c \d_x n - \d_x (D \d_x n) +\cdots \, ,
\end{equation}
with velocity $c=\nu/(2\pi \chi)$. Linearizing in the fluctuations $n = \bar n + \delta n$, { the standard hydrodynamic approach \cite{KADANOFF1963419} yields the retarded Green's function}
\begin{equation}\label{eq_G_naive}
G^R_{nn}(\omega,k)
	= \chi\frac{ick + D k^2}{-i(\omega-ck) + Dk^2} + \cdots\, ,
\end{equation}
where the corrections $\cdots$ are less singular as $\omega,k \to 0$. Here and in the following, functions of $n$ such as $c,\,D,\,\chi$ are evaluated on the background density $\bar n$. In the absence of an anomaly, the velocity $c$ vanishes and one obtains a diffusive Green's function as expected. The linear analysis suggests that the anomaly $\nu\neq 0$ leads to a right-moving ballistic front at velocity $c=\nu/(2\pi \chi)$, with diffusive spreading around the front \cite{PhysRevLett.72.4129,PhysRevB.51.13449}. We will see that this conclusion is incorrect. The chiral ballistic front is tied to the bulk Hall conductivity
\begin{equation}\label{eq_sigmaxy}
\sigma_{xy}^{\rm bulk}
	= \lim_{\omega \to 0}G^R_{j_xn}(\omega,0) = \chi c = \frac{\nu}{2\pi}\, ,
\end{equation}
and is a robust consequence of the anomaly. However, dissipation does not lead to diffusive spreading around the chiral front, because of a breakdown of the perturbative expansion in dissipative hydrodynamics. This can be seen by expanding the equation of motion \eqref{eq_eom} beyond leading order in $\delta n$ (which we denote as $n$ in the following for simplicity): { writing $c(n)\simeq c + c' n$ with $c'\equiv \d c / \d n = - \frac{\nu}{2\pi} \frac{\chi'}{\chi^2}$ and $\chi' \equiv \d\chi/\d n$, one finds}
\begin{equation}\label{eq_eom2}
\p_x\eta_x =
	\dot n + c \d_x n + \frac12 c' \d_x  n^2 - D \d_x^2  n +\cdots \,  .
\end{equation}
In the absence of additional symmetries, there is no reason for $c'$ to vanish and nonlinearities are generically expected, see e.g.~Ref.~\cite{PhysRevLett.108.206810}. Less relevant nonlinear terms coming from the $n$ dependence of the diffusivity $D$ are omitted. We included a noise term $\eta_x$ in the constitutive relation, whose symmetric Green's function is constrained by the fluctuation-dissipation theorem at leading order in gradients to be $\langle \eta_x(t,x)\eta_x\rangle = 2D\chi T \delta(x)\delta (t)+\cdots$. To establish the leading correction to ballistic propagation, it is convenient to work in the frame of the chiral front $x' = x-ct$, $t'=t$ (or, equivalently, $\omega'=\omega + ck$, $k'=k$). In these coordinates the equation of motion
\begin{equation}\label{eq_eom_boost}
\p_x\eta_x =
	\d_{t'} n + \frac12 c' \d_{x}  n^2 - D \d_{x}^2  n +\cdots \,  ,
\end{equation}
leads to a scaling $\omega' \sim k^2$, so that $\eta \sim k^{3/2}$ and $n\sim k^{1/2}$. One then finds that the interaction term $c'$ is {\em relevant}, and drives the system to a new dissipative fixed point that is not described by diffusive spreading around the chiral front \eqref{eq_G_naive}. In terms of the original coordinates, we expect $\omega -ck\sim  k^z$, with $z<2$ at the stable fixed point. In fact, Eq.~\eqref{eq_eom_boost} is nothing but the KPZ equation, with charge mapping to the slope of the interface $n = \d_x h$, and the system is described by Burgers--KPZ \cite{PhysRevA.16.732,PhysRevLett.56.889} universality with $z=3/2$. The symmetric Green's function is given by
\begin{equation}\label{eq_G_true}
G_{nn}(\omega,k)
	= \frac{\chi T}{\mathcal D k^{z}} g_{\rm KPZ} \left(\frac{\omega - ck}{\mathcal D  k^z}\right) + \cdots\, ,
\end{equation}
where $\cdots$ are terms that are subleading in the scaling $\frac{\omega - ck}{k^z}\sim1$, and $g_{\rm KPZ}$ is the KPZ scaling function which is known numerically to high precision \cite{Praehofer2004} (see appendix \ref{app_response} for some of its properties and  {our conventions for Green's functions}). $G_{nn}(\omega,k)$ is sharply peaked around $\omega = ck$ with a width of order $\mathcal D k^z$; charge fluctuations therefore obey the dispersion relation
\begin{equation}\label{eq_disp_rel}
\omega = c k - i \mathcal D k^z + \cdots\, .
\end{equation}
KPZ scaling ties dissipation to thermodynamics: the dimensionful constant $\mathcal D\sim {\rm length}^z/{\rm time}$ is fixed in terms of parameters in the equation of motion by dimensional analysis
\begin{equation}\label{eq_our_D}
\mathcal D
	= \sqrt{T \chi} |c'|
	= \sqrt{\frac{T}{\chi^3}} \frac{|\nu|}{2\pi} |\chi'|\, .
\end{equation}
This expression makes manifest three crucial ingredients that led to KPZ universality around the chiral front: finite temperature $T$, the anomaly $\nu$, and thermodynamic nonlinearities through $\chi' = \d \chi /\d n$.

Equation \eqref{eq_G_true} leads to a universal prediction for transport on the edge: the symmetric Green's function controls the dissipative optical conductivity at low frequencies $\omega \tau_{\rm th}\ll 1$ through the fluctuation-dissipation theorem and a Ward identity -- one finds
\begin{equation}
\sigma(\omega)
	\simeq\lim_{k\to 0} \frac{\chi \omega^2}{2\mathcal D k^{7/2}}  g_{\rm KPZ}\left(\frac{\omega}{\mathcal D k^{3/2}}\right)
	= a \frac{\chi \mathcal D^{4/3}}{\omega^{1/3}} \, ,
\end{equation}
with $a\approx 0.417816$ (see appendix \ref{app_response}).
{ While singular conductivites are common in one-dimensional momentum conserving systems \cite{PhysRevLett.89.200601,doi:10.1080/00018730802538522,PhysRevLett.122.206801}, momentum conservation was not assumed here.  This singularity }
as $\omega \to 0$ will be regulated in a system of finite length $L$, see Ref.~\cite{doi:10.1080/00018730802538522} for a discussion on subtleties with the Kubo formula in this situation. Although $\lim_{\omega\to 0}\sigma(\omega,k)$ vanishes for $k\neq 0$, the relevant observable may be $\sigma(\omega,k)$ at $\omega\sim ck\sim c/L$ \cite{PhysRevLett.89.200601,doi:10.1080/00018730802538522}, in which case one finds $\sigma_{\rm dc} \sim \chi \mathcal D^{4/3} (L/c)^{1/3}$. This also leads to a thermal contribution to the current noise $S_{\rm th}= \sigma_{\rm dc} / L \sim L^{-2/3}$, which vanishes more slowly than the standard thermal contribution $S_{\textrm{th}}\sim L^{-1}$. It would be interesting to explore the relevance of this correction in shot noise measurements in QH systems \cite{PhysRevB.51.2363,PhysRevLett.72.724,heiblum2019edge}.

Non-dissipative response such as the bulk Hall conductivity  $\sigma_{xy}^{\rm bulk}$ is controlled instead by the real part of the retarded Green's function $\Re G^R_{nn}$, which can be obtained from $\Im G^R_{nn}$ by analyticity. This is done appendix \ref{app_response}, where we show that the quantized bulk Hall conductivity  \eqref{eq_sigmaxy} is unchanged. Finally, long-range Coulomb interactions can be taken into account as usual in the random phase approximation by resumming a geometric series of diagrams involving photons -- this does not qualitatively change the dispersion relation, which simply receives logarithmic corrections, see e.g.~\cite{volkov1988edge}.

\section{The fate of heat}\label{sec_heat}

We now extend the discussion to include the other hydrodynamic mode: heat, or energy.
Heat has less privileged a status than charge, since it can leak out of the edge through phonons and will therefore only be approximately conserved. However, the time scale for heat loss may be parametrically longer than $\tau_{\rm th}$ as it is controlled by different physics -- a possibility affirmed by the experimental observation of the collective heat mode \cite{Bid2010}. Neglecting first thermoelectric effects, charge and heat decouple and can be treated separately. A nonzero bulk thermal Hall conductivity $\kappa_{xy}$ then gives heat a finite chiral speed of sound~\cite{PhysRevB.55.15832} $c_{\rm heat} = \frac{\kappa_{xy}}{c_V}$, where $c_V$ is the specific heat, and the analysis in section \ref{sec_main} holds with heat replacing charge. This result is largely unaffected by coupling between charge $n$ and energy $\varepsilon$ -- { expanding the continuity relations as in Eq.~\eqref{eq_eom2} now yields a system of KPZ equations}
\begin{equation}\label{eq_KPZ_matrix}
\d_t n_a + C_{ab} \nabla n_b + D_{ab} \nabla^2n_b  + \lambda_{abc} n_b \nabla n_c + \cdots= 0 \, ,
\end{equation}
with $n_1=n$ and $n_2=\varepsilon$. As long as the velocity eigenvalues are distinct, going into the rest frame of any eigenmode one finds that interactions with the other eigenmodes are kinematically disfavored. One therefore expects two independent KPZ modes around each chiral ballistic front -- this is indeed what is observed numerically %
	\footnote{The case where both velocity eigenvalues are equal is more subtle and has been explored in Ref.~\cite{PhysRevLett.69.929}.} \cite{Ferrari2013}.

One important exception is when the thermal Hall conductivity vanishes $\kappa_{xy} = 0$, so that the heat mode does not propagate ballistically \cite{PhysRevB.55.15832} (this happens e.g.~for $\nu=2/3$). Although a linearized analysis would suggest that heat then diffuses, its nonlinear coupling to the fluctuating charge mode also leads to a breakdown of diffusion in this case. This nonlinear coupling comes from the fact that in a background field $F_{0x} = E_x$, the energy continuity relation is changed to $\dot \varepsilon + \d_x j^{\varepsilon}_x = E_x j_x$,
which using \eqref{eq_consti} fixes the leading term in the constitutive relation for the energy current %
	\footnote{This equation was discussed in a relativistic context in \cite{Valle:2012em,Jensen:2012kj}, and a non-relativistic (but still translation invariant) context in \cite{Bradlyn:2014wla}; we stress that it is derived here using only the continuity relations for charge and energy.}
$
j^{\varepsilon}_x = \frac{\nu}{4\pi}\mu^2 +\cdots\, .
$
The two modes that diagonalize the $C$ matrix in \eqref{eq_KPZ_matrix} are now $\delta \mu = \chi^{-1}_{nn}\delta n + \chi^{-1}_{n\varepsilon} \delta \varepsilon$ and $\delta s = \frac{1}{T}(\delta \varepsilon - \mu \delta n)$. The former is still described by KPZ universality, with a correlator of the form \eqref{eq_G_true}. Instead when $\kappa_{xy}=0$, entropy fluctuations have a vanishing speed and self-coupling $ \lambda_{sss} = 0$. However coupling to the KPZ mode $\lambda_{s \mu \mu}\neq 0$ leads to superdiffusion $\omega_{\rm heat}\sim -i k^{z_{\rm heat}}$. Although the exponent is not known analytically, a ``mode coupling'' approximation gives $z_{\rm heat} = 5/3$ and seems consistent with numerics (see Refs.~\cite{doi:10.1080/00018730802538522,Spohn2014} for reviews). { This approximation yields again $\bar\kappa(\omega) \sim 1/\omega^{1/3}$; however soft heat modes now lead to a more singular charge conductivity $\sigma(\omega) \sim 1/\omega^{2/5}$. It is interesting that soft fluctuations are further enhanced in the $\nu=2/3$ state, where experiments have suggested higher sensitivity to finite system size \cite{Banerjee2017}.}

\section{Higher dimension: Surface chiral metals and chiral magnetic effect}\label{sec_surface}

Two reasons drive us to generalize the theory of Section \ref{sec_main} to higher dimensions: first, we will find that the KPZ fixed point can be accessed perturbatively from the upper critical dimension $d_c=2$; second, chiral systems with diffusive broadening naturally occur in higher dimensions as well.
In $d=2$ the theory we consider furnishes the low-energy description of surface chiral metals \cite{PhysRevLett.76.2782,PhysRevLett.75.4496}. These are boundaries of three-dimensional materials made from layered QH systems -- they exhibit propagation of a chiral diffusive front in the direction of the layer and regular diffusion in the transverse direction,
which was shown to be stable against localization \cite{PhysRevLett.76.2782}.
In $d=3$, the theory describes the hydrodynamics of a charge current subject to the chiral magnetic effect in the presence of a background magnetic field (decoupling momentum and energy fluctuations).
The chiral magnetic effect \cite{PhysRevD.22.3080} corresponds to a nonvanishing equilibrium value of the charge current in the presence of a magnetic field, and is due to the chiral anomaly. This effect arises in condensed matter systems such as Weyl semimetals \cite{Hosur_2013,PhysRevB.88.104412}, in heavy ion physics \cite{PhysRevD.78.074033} and astrophysics \cite{Charbonneau_2010}.

\begin{figure}
\centerline{\begin{overpic}[width=0.35\textwidth,tics=10]{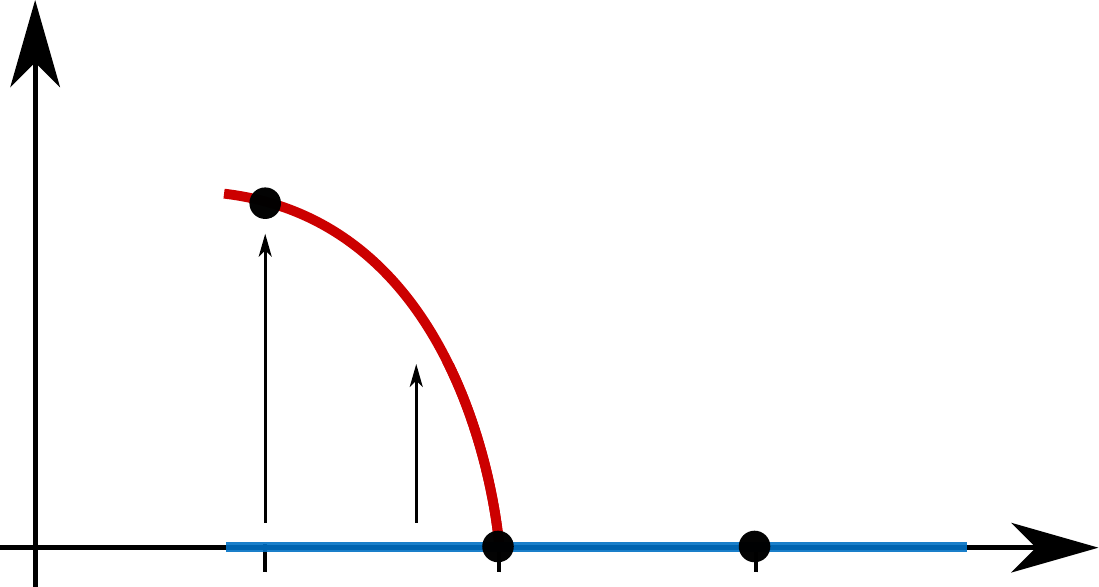}
	 \put (-7,47) {\large$\lambda$}
	 \put (104,2) {\large$d$}
	 \put (23,-4) {$1$}
	 \put (44,-4) {$2$}
	 \put (67.5,-4) {$3$}
	 \put (18,40) {{\color{inkred}\bf  KPZ}}
	 \put (50,10) {{\color{inkblue} \bf Chiral diffusion}}
\end{overpic}}
\caption{Fixed points $\lambda^*$ as a function of spatial dimension $d$.
\label{rg1}}
\end{figure}
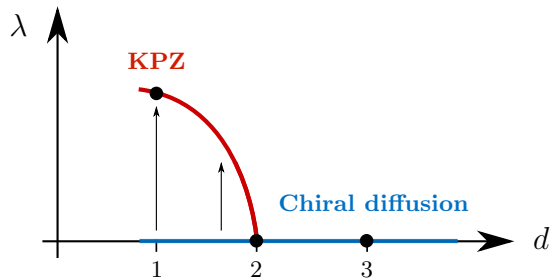

The common feature to all such systems is the presence of a chiral front with diffusive broadening along a given direction, which we label with $x$, and of ordinary diffusion in the orthogonal directions, which we label with $y^A$, where $A=2,\dots,d$. Up to first order in gradients the constitutive relations for the current are
\begin{equation}\label{curr}
j_x=\frac \nu{2\pi} \mu -\chi D_x \partial_x \mu,\qquad {j}_A=- \chi D_\perp \p_A \mu\ ,
\end{equation}
where the chemical potential $\mu$ is an arbitrary function of the charge density $n$.
Working again in the frame of the chiral front $x'=x-ct$, $y'=y$, $t'=t$, the conservation equation for Eq.~\eqref{curr} reads
\begin{equation}\label{conss1}
\p_{t'} n+\frac12 c'\p_x n^2-D_{ij} \partial_i \partial_jn=\p_i \eta^i+ \cdots \ ,
\end{equation}
where $D_{ij} = {\rm diag}(D_x,D_\perp, \ldots,D_\perp)$. The correlator of the noise current  $\eta^i=(\eta^x,\eta^A)$ is again fixed by thermal equilibrium: $\langle \eta_i(t,x,y) \eta_j\rangle= 2 D_{ij} \chi T \delta(t)\delta(x)\delta^{(d-1)}(y)$. In this frame, $\omega\sim k^2$, implying that $\eta^i\sim k^{\frac d2+1}$, $n\sim k^{\frac d2}$, and thus $c'$ scales as $k^{\frac{2-d}2}$, i.e. the interaction becomes marginal in $d=2$ and irrelevant in $d>2$. This stochastic system was first studied in Ref.~\cite{PhysRevLett.54.2026} in the context of driven diffusive systems, where the drive plays a crucial role in enhancing hydrodynamic fluctuations. We emphasize that our system is not driven:
instead the anomaly
enhances fluctuations. To study the effects of fluctuations and determine the RG fate in $d=2$ we implement the effective field theory approach to hydrodynamics \cite{Crossley:2015evo,Glorioso:2017fpd}, reviewed in appendix \ref{app_eft}. This framework allows us to perform a dynamical RG analysis keeping all the symmetries manifest, and appropriately capturing contact terms in correlation functions. The central object is the path integral
\be Z=\int Dn D\varphi_a e^{iS[n,\varphi_a]}\ ,\ee
where $\varphi_a(t,x,y^A)$ is an auxiliary field, and can be related to the noise currents in Eq.~\eqref{conss1} following the Martin-Siggia-Rose (MSR) formalism \cite{martin1973statistical}. The action associated to the stochastic equation (\ref{conss1}) is given in appendix \ref{app_eft}. Renormalization can be studied as usual by integrating out modes in a momentum shell $M\leq |\vec k|\leq \Lambda$. We find that $\chi,c',D_\perp$ do not renormalize. It is illuminating to express the renormalization of $D_x$ in terms of a rescaled coupling.
To this aim, we rescale $\p_x\to \p_x/\sqrt{D_x}$, $\p_A\to \p_A/\sqrt{D_\perp}$, $\varphi_a\to \varphi_a/\sqrt{T\chi}$, $n\to n\sqrt{T\chi}$ to canonically normalize the fields. Then the cubic coupling becomes $\lambda=c'\sqrt{\frac{\chi T}{D_x}}$. The $\beta$-function for $\lambda$ is
\begin{equation}\label{eq_rg}
\beta_{\lambda}=-\frac\vep 2 \lambda+\frac {\lambda^4}{32\pi c'\sqrt{D_\perp \chi T}}\ ,
\end{equation}
which shows that the coupling is marginally irrelevant in $d=2$. For $d=2-\varepsilon$, the diffusive fixed point with $\lambda=0$ is unstable, and the stable fixed point can be accessed perturbatively: $\lambda^{*3}=16\pi\vep c'\sqrt{D_\perp\chi T}$. As $d\to 1$, the fixed point will approach the KPZ universality class.
For $d\geq2$, $\lambda=0$ becomes stable and is the only fixed point \footnote{Note that $\lambda$ has the same sign as $c'$, so $\lambda=0$ is the only fixed point for $\vep<0$.}. This is summarized in Fig.~\ref{rg1}. It is interesting that the generalization to higher dimensions \eqref{conss1} is distinct from the one natural for the KPZ equation, where interactions are instead marginally relevant in $d=2$.

For surface chiral metals in $d=2$ where interactions are marginally irrelevant, the chiral diffusive fixed point is approached slowly and transport parameters run logarithmically. In this case one can solve the RG flow equation \eqref{eq_rg} and find the conductivity at low frequencies
\begin{equation}
\sigma_{xx}(\omega)=\chi D_x(\omega)\simeq \chi\left[\frac{3\chi T c^{'2}}{32\pi\sqrt{D_\perp}}\log \frac{1}{\omega}\right]^{2/3}\, .
\end{equation}
In $d=3$, pertaining to the chiral magnetic effect, the coupling $\lambda$ is irrelevant. As shown in appendix \ref{app_eft}, one finds $\sigma_{xx}(\omega)\sim \sigma_{xx}(0)+\lambda^2\omega^{\frac 12}$. This is the same frequency power generated by fluctuation corrections in three-dimensional momentum-conserving systems.

\section{Discussion}

We have shown that hydrodynamic fluctuations on edges realizing the chiral anomaly lead to a breakdown of diffusion, giving rise to singular low frequency transport and anomalous damping of edge modes. In particular, edge magnetoplasmons are predicted to be anomalously damped, Eqs.~(\ref{eq_disp_rel}, \ref{eq_our_D}). The charge susceptibility $\chi$ and $\chi'= \d \chi/\d n$ are non-universal but expected to have weak field and temperature dependence. The linear dependence of damping on filling $\nu$ for gapped bulks has been widely observed experimentally, see e.g. Refs.~\cite{PhysRevB.45.3894,PhysRevB.50.1582}. The temperature and wavevector dependence of damping have been less systematically reported -- the weak temperature dependence of damping observed in graphene in Ref.~\cite{PhysRevLett.113.266601} is consistent with Eq.~\eqref{eq_our_D}. To our knowledge only Ref.~\cite{PhysRevB.50.1582} studied wavevector dependence; the dependence they observe is between  linear to quadratic, which would neatly agree with Eq.~\eqref{eq_disp_rel}. Moreover, the overall size of damping observed is consistent with Eqs.~(\ref{eq_disp_rel}, \ref{eq_our_D}): estimating $\chi'\sim \omega_c$ to be of order the bulk gap and $\chi\sim 1/c$, one finds a quality factor $Q \sim \sqrt{\hbar \omega T}/\omega_c \sim 100$ at $\omega \sim 20 \hbox{ MHz}$, consistent with Ref.~\cite{PhysRevB.50.1582}. However more thorough investigation --  which is well within experimental reach -- is needed to unequivocally confirm our prediction.

In the presence of edge reconstruction \cite{PhysRevB.46.4026} with weak interedge interactions, our results can, in principle, apply to each edge (and perhaps most usefully to the outermost one), with the appropriate anomaly. If interedge interactions are strong enough, we instead expect only a single collective charge and heat mode to be long lived.

Nonlinear charge fluctuations have been argued to lead to a breakdown of the linearized edge picture even at $T=0$ in the absence of dissipation, where the Burger's equation is stabilized not by the diffusive term in \eqref{eq_eom} but by a non-dissipative two-derivative term (the Benjamin-Ono equation) \cite{PhysRevLett.108.206810}, whose coefficient is related to the bulk Hall viscosity. This term leads to a non-analytic correction to the dispersion relation $\omega -ck \sim k |k|$ -- we expect this non-analyticity softens at finite temperature, and is then less relevant than the diffusive term.

Although we have focused on charge and heat modes at the edge of QH systems, the necessary ingredients -- chiral edge modes protected by a continuity relation -- are realized in a number of other systems, where similar conclusions will hold for transport of charge, heat or spin on the boundary. These include gapped quantum spin liquids and topological superconductors \cite{Kasahara2018}, quantum anomalous Hall \cite{Konig766} and quantum spin Hall systems \cite{Chang167} (when spin conservation is a good approximation).

\subsection*{Acknowledgements}

We are grateful to Andrey Gromov, Steve Kivelson, Woowon Kang, Matthew Lapa, Shinsei Ryu, Dam T.~Son, Hassan Shapourian, John Toner, Ho-Ung Yee and Paul Wiegmann for inspiring discussions. LVD is supported by the Swiss National Science Foundation and a Robert R. McCormick Postdoctoral Fellowship. PG is supported by a Leo Kadanoff Fellowship.


\bibliography{chi_hyd}{}


\pagebreak
\appendix

\onecolumngrid

\section{Response functions}\label{app_response}

We work with two types of charge response functions in the main text -- the retarded Green's function and symmetric Greens function
\begin{equation}
G^R_{nn}(t,x)
	\equiv i\theta(t) \langle [n(x,t),n(0,0)]\rangle_\beta\, , \qquad
G_{nn}(t,x)
	\equiv \frac{1}{2} \langle \{n(x,t),n(0,0)\}\rangle_\beta\, ,
\end{equation}
where $\langle \cdot \rangle_\beta \equiv \Tr (e^{-\beta H} \cdot) / \Tr (e^{-\beta H} )$. Their Fourier transforms are related by the fluctuation-dissipation theorem: in the hydrodynamic limit ($\omega \ll \frac{1}{\tau_{\rm th}}\lesssim T$)
\begin{equation}\label{eq_app_fdt}
\Im G^R_{nn}(\omega,k)
	= \tanh \frac{\omega}{2 T} G_{nn}(\omega,k)
	\simeq  \frac{2 T}{\omega} G_{nn}(\omega,k)\, .
\end{equation}
%

\subsection{Conventions for the KPZ scaling function}

Charge density on the edge maps to the KPZ slope $n \leftrightarrow \d_x h$. In this appendix, we will translate between our conventions for the scaling function $g_{\rm KPZ}$ in Eq.~\eqref{eq_G_true} and those in Ref.~\cite{Praehofer2004}, where the scaling function was studied to high precision. From the equation of motion in the `pulse frame'
\begin{equation}
\eta =
	\dot n + \frac12 c' \d_x  n^2 - D \d_x^2  n +\cdots \,  ,
\end{equation}
we see that the standard KPZ coupling \cite{PhysRevLett.56.889,Praehofer2004} is $\lambda = -c' = \frac{\nu}{2\pi} \frac{\chi'}{\chi^2}$. Furthermore, the sum rule relates our susceptibility $\chi T$ to Eq.~(1.4) of \cite{Praehofer2004} as $A = \chi T$. The dimensionful constant chosen there (in Eq.~(1.6) of \cite{Praehofer2004}) is then related to ours \eqref{eq_our_D} as
\begin{equation}
\widetilde{\mathcal D}
	= \sqrt{2 \lambda^2 A}
	= \sqrt{2} \sqrt{\frac{T \chi'^2}{\chi^3}} \frac{|\nu|}{2\pi}
	= \sqrt{2} \mathcal D\, .
\end{equation}
The symmetric Green's function ($S(t,x)$ in the notation of \cite{Praehofer2004}) is related to the scaling function $f$ (Eq.~(1.8) in \cite{Praehofer2004}) as
\begin{equation}
G_{nn}(t,x)
	= S(t,x) = \frac{\chi T}{(\widetilde{\mathcal D} t)^{2/3}} f \left( \frac{x}{(\widetilde{\mathcal D} t)^{2/3}}\right)\, .
\end{equation}
This function has the following asymptotics
\begin{equation}
f(0) = \hbox{const}\, , \qquad\qquad
\lim_{y\to \infty} f(y) \approx e^{-{\rm const}'y^3}\, .
\end{equation}
Fourier transforming in space gives
\begin{equation}
G_{nn}(t,k)
	\equiv \int dx \, G_{nn}(t,x) e^{-ikx}
	= \chi T \hat f \left( k (\widetilde{\mathcal D} t)^{2/3}\right)\, ,
\end{equation}
where the scaling function $\hat f$ is defined in Eq.~(5.3) of \cite{Praehofer2004}. It has the properties
\begin{equation}\label{eq_PS_hatf}
\hat f(0) = 1 \quad \hbox{(sum rule)}\, , \qquad\qquad
\lim_{\kappa\to \infty}\hat f(\kappa)
	\approx {\rm const}\cdot \kappa^{-9/4} e^{-\frac{1+i}{2}\kappa^{3/2}} + \hbox{c.c.}\, .
\end{equation}
Finally, let us perform the time Fourier transform:
\begin{equation}
G_{nn}(\omega,k)
	= \int dt\, e^{i\omega t}G_{nn}(t,k)
	= \frac{\chi T}{\widetilde{\mathcal D} k^{3/2} } \stackrel{\circ}{f} \left( \frac{\omega}{\widetilde{\mathcal D} k^{3/2}}\right)\, ,
\end{equation}
where $\stackrel{\circ}{f}$ is the scaling function defined in Eq.~(5.9) of \cite{Praehofer2004}. Now given the definition of our scaling function $g_{\rm KPZ}$ in Eq.~\eqref{eq_G_true}, we see that our scaling function is related to $\stackrel{\circ}{f}$ by
\begin{equation}
g_{\rm KPZ}(w)
	= \frac{1}{\sqrt{2}} \stackrel{\circ}{f} \left(\frac{w}{\sqrt{2}}\right)\, .
\end{equation}
Eq.~(5.11) in \cite{Praehofer2004} implies that it has the properties
\begin{equation}\label{eq_kpz_limits_app}
\int \frac{d w}{2\pi } g_{\rm KPZ}(w) = 1\, , \qquad\quad
\lim_{w\to \infty}g_{\rm KPZ}(w) = \frac{2 a}{w^{7/3}}\, , \qquad\quad
g_{\rm KPZ}(0) = b\, ,
\end{equation}
with $a\approx 0.417816$ and  $b\approx 3.43730$. The first property guarantees that Eq.~\eqref{eq_G_true} satisfies the usual sum rule $\int \frac{d\omega}{2\pi} \, G_{nn}(\omega,k) = \chi T$.

Returning to the lab frame, the fluctuation-dissipation theorem \eqref{eq_app_fdt} relates the symmetric correlator to the retarded Green's function
\begin{equation}\label{eq_ImGR}
\Im G^R_{nn}(\omega,k)
	\simeq \frac{\omega}{2T} G_{nn}(\omega,k)
	\simeq \frac{\chi \omega}{2 \mathcal D k^{3/2}} g_{\rm KPZ} \left(\frac{\omega-ck}{\mathcal D k^{3/2}}\right)\, ,
\end{equation}
(in the hydrodynamic regime $\omega \ll T$). The dissipative optical conductivity can then be obtained from a Ward identity
\begin{equation}
\sigma(\omega,k)
	= \frac{1}{\omega} \Im G^R_{jj}(\omega)
	= \frac{\omega}{k^2} \Im G^R_{nn}(\omega)
	\simeq \frac{\chi \omega^2}{2 \mathcal D k^{7/2}} g_{\rm KPZ} \left(\frac{\omega-ck}{\mathcal D k^{3/2}}\right)\, .
\end{equation}
It is striking that response functions at the dissipative fixed point involve bare parameters from the equation of motion \eqref{eq_eom2}. This is possible because $\nu,\, \chi$ and $\chi'$ do not get renormalized \cite{PhysRevA.16.732}, as discussed in below.

\subsection{Retarded response}

Analyticity of retarded response $G^R(\omega)$ in the upper-half complex plane implies that its real and imaginary parts are connected by the Kramers-Kronig relation (see e.g. \cite{coleman2015introduction}). The dynamic susceptibility can be obtained from Eq.~\eqref{eq_ImGR} as
\begin{equation}
\chi(z,k)
	= \int \frac{d\omega}{\pi} \frac{\Im G^R_{nn}(\omega,k)}{\omega-z}
	\simeq \chi + \chi \frac{z}{\mathcal D k^{3/2}} \int \frac{dx}{2\pi} \frac{g_{\rm KPZ} (x)}{x - \frac{z - ck}{\mathcal D k^{3/2}}}\, .
\end{equation}
The full retarded Green's function is then the analytic continuation of $\chi(z,k)$ to the upper-half plane
\begin{equation}
G^R_{nn}(\omega,k)
	= \chi(\omega + i 0^+,k)
	\simeq \chi + \chi \frac{\omega}{\mathcal D k^{3/2}} \int \frac{dx}{2\pi} \frac{g_{\rm KPZ} (x)}{x - \frac{\omega - ck}{\mathcal D k^{3/2}}-i0^+}\, .
\end{equation}
Non-dissipative response can now be studied from the real part of $G^R_{nn}$. For example the bulk Hall conductivity is as expected
\begin{equation}
\sigma_{xy}^{\rm bulk}
	= \lim_{\omega \to 0}\lim_{k\to 0} G^R_{j n}(\omega,k)
	= -\lim_{\omega\to 0}\lim_{k\to 0} \frac{\omega}{k} G^R_{nn}(\omega,k)
	= \chi c \int \frac{dx}{2\pi} g_{\rm KPZ}(x) = \frac{\nu}{2\pi}\, ,
\end{equation}
where in the last step we used \eqref{eq_kpz_limits_app}. Note that use of the Ward identity can introduce contact terms, which we are not keeping track of here. These analytic contributions to transport can be controlled if desired with the effective field theory framework presented in appendix \ref{app_eft}.


\section{Renormalization of chiral diffusion}\label{app_eft}

\subsection{Effective field theory of hydrodynamics}

Hydrodynamics has recently been cast into an effective field theory (EFT) framework which is based solely on symmetry principles \cite{Crossley:2015evo,Glorioso:2017fpd,Haehl:2015foa,Jensen:2017kzi}.
Here we briefly review this framework, a more complete exposition can be found in \cite{Glorioso:2018wxw}.
In the next subsections we will apply the formalism to study the dynamical renormalization group of chiral diffusion. For the systems studied in the main text, the effective action one obtains essentially coincides with that obtained from the stochastic equation (\ref{conss1}) using the MSR formalism \cite{martin1973statistical}. The strength of the EFT approach is that it guarantees that all terms associated with the chiral anomaly and fluctuations are captured in a systematic derivative expansion of the action, including contact terms, and it is straightforwardly generalized to more complicated systems. A systematic study of renormalization in the context of diffusion was undertaken in \cite{Chen-Lin:2018kfl}.

We are interested in studying the low-energy behavior of real-time correlation functions of the charge and current operators $j^0(t,x),j^i(t,x)$ at finite temperature.
These correlation functions are conveniently encoded in the Schwinger-Keldysh generating functional. This is obtained by coupling the system to a background gauge field $A_\mu$ as $S\to S+\int j^\mu A_\mu$, where $S$ is the action of the underlying microscopic system, and $\mu=0,i$ is a spacetime index. One then writes the generating functional
\be \label{gen1}Z[A_{1\mu},A_{2\mu}]\equiv \Tr\left(U(A_{1\mu})\rho_0 U^\dag(A_{2\mu})\right)\ ,\ee
where $U(A_{\mu})$ is the evolution operator from $t=-\infty$ to $t=+\infty$ for the action $S+\int J^\mu A_\mu$, and $\rho_0=\frac{e^{- H/T}}{\Tr e^{- H/T}}$ is the thermal density matrix for the Hamiltonian $H$ at temperature $T$ describing the state of the system at the initial time $t=-\infty$. Varying $Z[A_{1\mu},A_{2\mu}]$ with respect to $A_{1\mu}$ ($A_{2\mu})$ introduces time-ordered (anti time-ordered) insertions of $j^\mu$ in the trace, which we denote by $j_1^\mu$ ($j_2^\mu$), i.e.:
\be \langle j^\mu_1(t,\vec x) j^\nu_2(t',\vec x')j_1^\rho(t'',\vec x'')\cdots\rangle\equiv \Tr\left({\mathcal T}(j^\mu(t,\vec x)j^\rho(t'',\vec x'')\cdots)\rho_0\tilde {\mathcal T}(j^\nu(t',\vec x')\cdots)\right)\ ,\ee
where ${\mathcal T}$ and $\tilde {\mathcal T}$ denote time- and anti time-ordering, respectively. This generating functional satisfies the following important constraints:
\begin{eqnarray}\label{norm}
\text{normalization:}\quad && Z[A_\mu,A_\mu]=1\\
\text{reflectivity:}\quad && Z[A_{1\mu},A_{2\mu}]=(Z[A_{2\mu},A_{1\mu}])^*\\
\text{KMS invariance:}\quad && Z[A_{1\mu},A_{2\mu}]=Z[A_{1\mu}(-t,-x,y^A),A_{2\mu}(-t-i T^{-1},-x,y^A)]\label{kms}\\
\text{gauge invariance:}\quad &&Z[A_{1\mu},A_{2\mu}]=Z[A_{1\mu}+\p_\mu \lambda_1,A_{2\mu}+\p_\mu\lambda_2]\label{gaugesk}
\end{eqnarray}
where $\lambda_1,\lambda_2$ are arbitrary functions of $t,x$. The first three properties are general for any Schwinger-Keldysh generating functional with $\rho_0$ being thermal at temperature $T$, where for the third property we additionally assumed the microscopic Hamiltonian $H$ to be invariant under the product of time-reversal and reflection in the $x$-direction, as is the case for the systems considered in the main text. The fourth property is a consequence of conservation of the currents $j_1^\mu$ and $j_2^\mu$.

The existence of long-lived hydrodynamic modes makes the generating functional (\ref{gen1}) non-local. These modes arise due to the conservation of $j_1^\mu$ and $j_2^\mu$. The crucial step is then to integrate back in these degrees of freedom \cite{Crossley:2015evo}, which leads to
\be\label{pi} Z[A_{1\mu},A_{2\mu}]=\int D\varphi_1 D\varphi_2 e^{iS[B_{1\mu},B_{2\mu}]}\ ,\ee
where $S[B_{1\mu},B_{2\mu}]$ is a \emph{local} functional of $B_{1\mu}=A_{1\mu}-\p_\mu \varphi_1$ and $B_{2\mu}=A_{2\mu}-\p_\mu \varphi_2$, and $\varphi_1,\varphi_2$ are the modes associated to the conservation of $j_1^\mu,j_2^\mu$, respectively. The dependence of the effective action $S$ on these combinations guarantees that, integrating out $\varphi_1,\varphi_2$, one obtains a gauge-invariant $Z$, i.e. the equations of motion for $\varphi_1,\varphi_2$ are precisely the conservations of $j_1^\mu,j_2^\mu$. Locality of $S$ follows from that, in our case, $\varphi_1,\varphi_2$ are the only long-lived modes in the system. The action $S$ should satisfy the same constraints as $Z$, i.e. eqs. (\ref{norm})-(\ref{gaugesk}) in terms of $B_{1\mu},B_{2\mu}$ instead of $A_{1\mu},A_{2\mu}$. Following the effective field theory approach, one writes down the most general terms for $S$ according to derivative expansion. It will be convenient to introduce a new basis of fields:
\be \varphi_r=\frac 12(\varphi_1+\varphi_2),\qquad \varphi_a=\varphi_1-\varphi_2\ .\ee
We also define $A_{r\mu}=\frac 12(A_{1\mu}+A_{2\mu})$ and $A_{a\mu}=A_{1\mu}-A_{2\mu}$, and similarly we introduce $B_{r\mu},B_{a\mu}$. As we will see, $\dot\varphi_r$ will be identified with $\mu$ in (\ref{curr}), while $\varphi_a$ will be related to the noise.  Since in diffusion $\mu$ is the only combination appearing in the equations, we impose the further symmetry on the action:
\be\label{diags} \varphi_r\to \varphi_r+\chi(x)\ ,\ee
i.e. the action should be invariant under time-independent shifts of $\varphi_r$, which allows $\dot \varphi_r$ to appear explicitly, but not $\p_i \varphi_r$.

\subsection{Action for chiral diffusion}
The general approach is to write down all terms compatible with the symmetries and constraints discussed in the previous subsection, order by order in derivatives. 
We first write down the non-chiral part of the action, which is separately invariant under spatial inversion in the $x$- and $y^A$-directions. The action is
\be
\label{s0}S_0=\int dt d^dx \left(n (B_{r0})B_{a0}-\sigma_x \dot B_{rx}B_{ax}-\sigma_\perp \dot B_{rA}B_{aA}+i\sigma_x TB_{ax}^2+i\sigma_\perp TB_{aA}^2 + \cdots\right)\ ,
\ee
where $\cdots$ denotes higher derivative terms. $n=n(B_{r0})$ is an arbitrary function of $B_{r0}$, and $\sigma_x,\sigma_\perp$ are constants. Note that we imposed rotation invariance in the $y^A$-directions.

We now add to $S_0$ the chiral contribution along the $x$-direction. Besides breaking the inversion symmetry $x\to -x$, the existence of these chiral hydrodynamic modes is tied to the presence of a chiral anomaly in the $x$-direction, $\p_\mu j^\mu=C F_{0x}$. For $d=1$ this was discussed in Section \ref{sec_main}. In $d=2$, for surface chiral metals, one can immediately generalize the $d=1$ case by considering stacking together quantum Hall layers in the limit of vanishing interlayer coupling, leading to the expected anomaly. Turning on interlayer couplings, as far as the bulk remains gapped, one expects that the anomaly inflow from the three-dimensional bulk to the boundary surface will guarantee that the anomaly is unmodified. For hydrodynamics in $d=3$ with chiral anomaly $\p_\mu j^\mu\propto F_{\mu\nu}\tilde F^{\mu\nu}$, a background magnetic field parallel to the $x$-direction $\vec B=B_0\hat x$ will lead to $\p_\mu j^\mu\propto F_{\mu\nu}\tilde F^{\mu\nu}= C F_{0x}$. To add the effect of the anomaly to the action we proceed in a way similar to \cite{Glorioso:2017lcn}. Eq. (\ref{gaugesk}) is modified to
\be \label{anom1}Z[A_{1\mu}+\p_\mu \lambda_1,A_{2\mu}+\p_\mu\lambda_2]=Z[A_{1\mu},A_{2\mu}]
e^{i\mathcal A}\ ,\ee
where the anomaly is
\be \mathcal A=C\int dt d^dx \left[\vep^{\alpha\beta}F_{\alpha\beta}^1\lambda_1
-\vep^{\alpha\beta}F_{\alpha\beta}^2\lambda_2\right]\ ,\ee
where $\alpha,\beta=t,x$, and $F^1_{\alpha\beta}=\p_\alpha A_{1\beta}-\p_\beta A_{1\alpha}$, and similarly for $F_{\mu\nu}^2$, where $\vep^{\alpha\beta}F_{\alpha\beta}=E_x$ for each of the two copies (we use the convention $\vep^{0x}=-1$). We add to $S_0$ a local action $S_{\text{ch}}$ which leads to the anomaly (\ref{anom1}). Under gauge transformations $\delta A_{1\mu}=\p_\mu\lambda_1$ and $\delta A_{2\mu}=\p_\mu\lambda_2$, we must have
\be\label{anS} \delta S_{\text{ch}}=C\int dt d^dx \left[\vep^{\alpha\beta}F_{\alpha\beta}^1\lambda_1
-\vep^{\alpha\beta}F_{\alpha\beta}^2\lambda_2\right]\ .\ee
A minimal action that satisfies this is
\be S_{\text{ch}}=C\int dtd^d x\vep^{\alpha\beta}(\varphi_1 F_{1\alpha\beta}-\varphi_2 F_{2\alpha\beta})\ .\ee
This action breaks (\ref{diags}):
\be \delta_\chi S_{\text{an}}=-2C\int dtd^d x\p_x \chi A_{a0}\ .\ee
To restore (\ref{diags}), we need to add another term to $S_{\text{ch}}$ which is gauge invariant (so that (\ref{anS}) is still satisfied) and makes $\delta_\chi S_{\text{ch}}=0$. The simplest choice is $S_{\text{ch}}\to S_{\text{ch}}+S_1$, where
\be S_1=2C\int dtd^d x(B_{1x} B_{10}-B_{2x}B_{20})\ .\ee
Written in terms of $r/a$ variables, we then have
\be S_{\text{ch}}=C\int dt d^d x\left(\varphi_a \vep^{\alpha\beta}F_{r\alpha\beta}+\varphi_r\vep^{\alpha\beta}F_{a\alpha\beta}+2 B_{ax}B_{r0}+2B_{rx}B_{a0}\right)\ .\ee
One can verify that constraints (\ref{norm})-(\ref{kms}) applied to $S_{\text{ch}}$ are satisfied. The complete minimal action for chiral diffusion is then $S=S_0+S_{\text{ch}}$.

Introduce
\be \tilde j_1^\mu=\frac{\delta S}{\delta A_{1\mu}},\qquad  \tilde j_2^\mu=\frac{\delta S}{\delta A_{2\mu}}\ .\ee
Note that these are the consistent currents, i.e. currents coming from varying the generating functional with respect to the background. Due to (\ref{anom1}), these currents are not gauge invariant \cite{Jensen:2012kj}. The currents can be made gauge invariant by shifting them with the Bardeen-Zumino term
\be j_{1,2}^\mu=\tilde j_{1,2}^\mu+2C\varepsilon^{\mu\nu}A_\nu\ ,\ee
where $j_1^\mu$ and $j_2^\mu$ are called covariant currents. In the main text, and in the following discussion, we shall always use covariant currents. Now, defining
\be  j_r^\mu=\frac 12( j_1^\mu+ j_2^\mu),\quad  j_a^\mu= j_1^\mu- j_2^\mu\ ,\ee
we find the following explicit expressions
\begin{gather}\label{currs}
j_r^0=n,\quad j_r^x=-\sigma\p_0 B_{rx}+2i\sigma_x TB_{ax}+4CB_{r0}\\
j_r^A=-\sigma\p_0 B_{rA}+2i\sigma_\perp T B_{aA},\quad j_a^0=\frac{\p n}{\p B_{r0}} B_{a0}\\
j_{a}^x=\sigma_x\p_0 B_{ax}+4CB_{a0},\quad j_a^A=\sigma_\perp\p_0 B_{aA}\ ,
\end{gather}
which are indeed gauge invariant. The $a$-type fields, as we will see, are related to the noise, while $j_r^\mu$ plays the role of $j^\mu$ of the main text. The same correspondence holds in the MSR formalism. Setting $a$-type fields to zero, as well as the background sources $A_{1\mu},A_{2\mu}$, the current reads
\be j_r^0=n,\quad j_r^x=4a\mu-\sigma\p_x\mu,\quad j_r^A=-\sigma\p_A\mu\ ,\ee
where we used the identification $B_{r0}=\p_0\varphi_r=\mu$, leading to the correct hydrodynamic constitutive relation of the current (\ref{curr}) upon setting $C=\frac \nu{8\pi}$.

We shall now rewrite the action in a slightly different form, which will make the physics more transparent. Note that, switching off the background sources $A_{1\mu},A_{2\mu}$, the action depends on $\varphi_r$ only through $\p_0\varphi_r=\mu$, due to (\ref{diags}). We can then make the change of variable to the path integral measure: $\int D\varphi_r \cdots=\int D\mu\cdots$, where we discarded the Jacobian of the transformation, $|\frac 1{\p_0}|$, being independent of all fields. We make the further change of variable $\int D\mu\cdots=\int D n J \cdots$, where $J=\det(\frac{\p n}{\p\mu}(t,\vec x)\delta^{(d)}(\vec x-\vec y)\delta(t-t'))$ is the Jacobian of the change of variable. In what follows, we shall use dimensional regularization, in which case the Jacobian factor in this transformation equals the identity (see e.g.~\cite{henneaux1994quantization}, Section 18.2.4). The path integral then becomes
\be Z=\int Dn D\varphi_a e^{iS[n,\varphi_a]}\ ,\ee
where
\be \label{actnf0}\begin{split}
S[n,\varphi_a]
	=&\int dt d^d x\left(-\varphi_a(\p_0 n+\p_x j_r^x+\p_A j_r^A)+i\sigma_x T (\p_x\varphi_a)^2+i\sigma_\perp T(\p_A\varphi_a)^2\right)\\
	=&\int dt d^d x\, \big(-\varphi_a(\dot n+c\p_xn+c'n\p_x n-D_x\p_x^2n-D_\perp \p_A^2 n)\\
	&\qquad\qquad+i\chi D_x T (\p_x\varphi_a)^2+i\chi D_\perp T(\p_A\varphi_a)^2\big)
\end{split}\,\ee
with $c=\frac {4C}\chi$, and $c'=-4C\frac{\chi'}{\chi^2}$, with $\chi'=-\chi^2\p_n^2\mu$. In the above, we also introduced the constants $D_x=\sigma_x/\chi$ and $D_\perp=\sigma_\perp/\chi$. Scaling $\omega\sim k$ to leading order, one finds that $\varphi_a\sim k^{\frac{d-1}{2}}$ and $n\sim k^{\frac{d+1}{2}}$, so that both the diffusive term $D$ and nonlinearity $c'$ are irrelevant corrections to the chiral ballistic front.

Following the discussion around Eq.~(\ref{conss1}), we make the change of coordinates $t'= t$, $x'= x-ct$, $y^{'A}= y^A$, for which the explicit action becomes (dropping the primes)
\be \label{actnf}S=\int dt d^d x\left(-\varphi_a(\dot n+c'n\p_x n-D_x\p_x^2n-D_\perp \p_A^2 n)+i\chi D_x T (\p_x\varphi_a)^2+i\chi D_\perp T(\p_A\varphi_a)^2\right)\ .\ee
Using steps similar to those of the MSR formalism, this path integral can be immediately shown to be equivalent to (\ref{conss1}).

\subsection{Renormalization}\label{ssec_renorm}

We will now evaluate the one-loop correction to the retarded two-point function of the charge, from which we will extract renormalization of transport. This can be written as the two-point function of $j_r^0$ and $j_a^0$ \cite{Glorioso:2018wxw} :
\be G^{R}_{nn}(t,x,y^A)=i\langle j_r^0(t, x,y^A) j_a^0(0)\rangle\ .\ee
In what follows we will neglect contact term contributions. Our computations will be based on the two-point functions of the degrees of freedom $n$ and $\varphi_a$ which, at the level of the quadratic part of the action (\ref{actnf}), are (using the notation $\left<\mathcal O_1\mathcal O_2\right>(p)=\int dt d^d x\,  e^{i\omega t-i\vec k \cdot \vec x}\left<\mathcal O_1(t,\vec x)\mathcal O_2(0)\right>$ )
\begin{eqnarray}
\label{nf}\langle n\varphi_a\rangle_0(p)&=&\frac{i}{i\omega-D_xk_x^2-D_\perp k_\perp^2}\\
\langle nn\rangle_0(p)&=&\frac{\chi T (D_xk_x^2+D_\perp k_\perp^2)}{|i\omega-D_xk_x^2- D_\perp k_\perp^2|^2}
\ ,
\end{eqnarray}
where $p^\mu=(\omega,\vec k)=(\omega,k^x,k_\perp^A)$, and $\langle n\varphi_a\rangle_0(p)$ is the tree-level part of $\langle n(p)\varphi_a(-p)\rangle$, and similarly for $\langle nn\rangle_0(p)$. Recall that we are in the coordinate frame following the chiral front. In the lab frame, the frequency is shifted to $\omega\to \omega+ck_x$. Notice that, from (\ref{currs}), the retarded two-point function of the charge density can be related to that of $n$ and $\varphi_a$:
\be i\langle J_r^0(p)J_a^0(-p)\rangle=-\chi\omega\langle n(p)\varphi_a(-p)\rangle\ ,\ee
where we neglected the non-linear term in $J_a^0=\chi\p_0\varphi_a+\chi' n \p_0\varphi_a$ as this is subleading at low energy. This can be inferred from the scaling analysis below Eq. (\ref{conss1}).

\begin{figure}[h]
\centerline{\begin{overpic}[width=0.28\textwidth,tics=10]{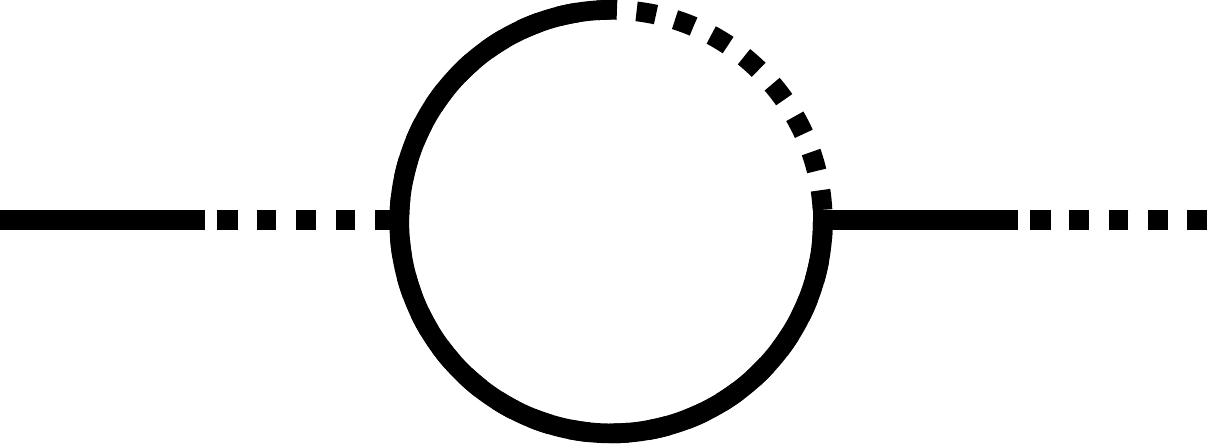}
	 \put (72,25) {\large$c'$}
	 \put (23,25) {\large$c'$}
\end{overpic}}
\caption{One-loop contribution to $\langle n\varphi_a\rangle$. Solid lines represent $n$ and dashed lines represent $\varphi_a$.
\label{loop}}
\end{figure}

We are now ready to compute loop corrections to (\ref{nf}). To obtain this we expand the path integral (\ref{pi}) in the interaction coupling $c'$ and perform Wick contractions
\be \langle  n \varphi_a\rangle = \langle  n \varphi_a\rangle_0+i\langle  n \varphi_aS_{\text{ing}}\rangle_0-\frac 12\langle  n \varphi_aS_{\text{ing}}^2\rangle_0 + \cdots \ ,\ee
where $S_{\text{int}}=c'\int dt d^d x\,n\p_x n$. For this simple action there is only one diagram contributing, shown in Fig. \ref{loop}. Loops of the form
$\int d^{d+1}p'\left<n\varphi_a\right>_0(p') \left<n\varphi_a\right>_0(p'-p)$
vanish as the propagator
$ \left<n\varphi_a\right>_0(p')$
is analytic in the upper frequency plane \cite{Gao:2018bxz}.

We represent this contribution as
\be  i\langle  n \varphi_a\rangle_0(p)\Sigma(p) \langle  n \varphi_a\rangle_0(p)\ .\ee
Higher loop contributions include a geometric series of diagrams like Fig.~\ref{loop} -- these are enhanced close to the pole and should therefore be resummed:
\be \label{twop}\langle n(p)\varphi_a(-p)\rangle=\frac{i}{i\omega-D_xk_x^2-D_\perp k_\perp^2+\Sigma(\omega,k_x,k_A)}\ ,\ee
where the self-energy is given by
\be\begin{split} -i\Sigma(\omega,k_x,k_A)&=c^{'2} k_x\int \frac{d^dk'}{(2\pi)^d}\int \frac{d\omega'}{2\pi}k_x'\langle nn\rangle(p-p')\langle n\varphi_a\rangle(p')\\
&=-i\chi T c^{'2}\frac{k_x^2}2\frac 1{\sqrt{2^d D_x D_\perp^{d-1}}}\int\frac{d^d q}{(2\pi)^d}\frac 1{\left(-i\omega+\frac12(D_xk_x^2+D_\perp k_A^2)\right)+q_x^2+ q_A^2}\ ,\end{split}\ee
where in the last step we performed the frequency integral, and changed to variables $(k'_x,k'_A)=\left(\frac{q_x}{\sqrt{2D_x}},\frac{q_A}{\sqrt{2D_\perp}}\right)+\frac 12 (k_x,k_A)$. In dimensional regularization,
\be \label{eq_dimreg} L_d(A)\equiv\int \frac {d^d u}{(2\pi)^d}\frac 1{A+u^2}=\frac{(A/\pi)^{\frac d2-1}}{2^d \Gamma\left(\frac d2\right)\sin\left(\frac{d\pi}2\right)}\ ,\ee
which diverges in even spatial dimensions (in dimensional regularization, power divergences vanish). From the scaling argument below Eq.~(\ref{conss1}) we already expect that the critical dimension is $d=2$, so we shall take $d=2-\varepsilon$ with $\varepsilon$ small, for which
\be L_{2-\varepsilon}(A)=\frac 1{2\pi\varepsilon}-\frac 1{4\pi}\left(\gamma+\log\left(\frac A{4\pi}\right)\right)+O(\varepsilon)\ ,\ee
where $\gamma$ is the Euler-Mascheroni constant, and the momentum cutoff $\Lambda$ is determined by $\frac 1\varepsilon=\log\Lambda$. We then find, in $d=2$,
\be \label{sigma0}\Sigma(\omega,k_x,k_A)=-\frac{\chi Tc^{'2} k_x^2}{16\pi\sqrt{D_x D_\perp}}\log\left(-i\omega+\frac{D_xk_x^2}2+\frac{D_\perp k^2_A}2\right)\ ,\ee
where we dropped analytic terms that can be absorbed in the diffusion constants. The diffusion constant $D_x$ receives a UV divergent contribution
\be\label{Dx} D_x^{(\Lambda)}=D_x+\frac{\chi Tc^{'2}}{8\pi\sqrt{D_xD_\perp}}\log\Lambda\ ,\ee
while $D_\perp$ does not renormalize at this order.

Let us now derive the $\beta$ functions of various couplings in dimension $d=2-\varepsilon$. Integrating out modes with $M<|k|<\Lambda$, Eq. (\ref{Dx}) leads to
\be \label{betad}\beta_{D_x}=\frac{\p D_x}{\p\log M}=\varepsilon D_x-\frac{\chi Tc^{'2}}{8\pi\sqrt{D_xD_\perp}}\ .\ee

From the form of (\ref{twop}) we see that there is no wavefunction renormalization, i.e. the first term of (\ref{actnf}) remains equal to $-1$. Also, going back to the lab frame $\omega\to \omega+ck_x$ it is obvious that $c$ in (\ref{actnf0}) does not renormalize, as $\Sigma\propto k_x^2$. Finally, we note that the action (\ref{actnf}) (or (\ref{actnf0})) has an emergent symmetry:
\be n \to n+\epsilon,\qquad x\to x+2\epsilon c' t\ ,\ee
which implies that $c'$ does not renormalize, thanks to absence of wavefunction renormalization. This can also be verified by directly compute renormalization of $c'$. In conclusion, at one loop, only $D_x$ undergoes renormalization.

Solving the RG flow equation (\ref{betad}) in $d=2$ gives, at leading order as $M\to 0$,
\be D_x(M)=\left(-\frac{3\chi T c^{'2}}{16\pi\sqrt{D_\perp}}\log M\right)^{2/3}\ .\ee
Now, at leading order the dispersion relation is $\omega\sim k^2\sim M^2$, and thus we find, for the conductivity of the system,
\be \sigma_{xx}(\omega)=\chi D_x(\omega)\to \chi\left(-\frac{3\chi T c^{'2}}{32\pi\sqrt{D_\perp}}\log\omega\right)^{2/3}\ .\ee
For nonvanishing $\omega,k_x,k_y$, one can fix $M$ by comparing to (\ref{sigma0}), which gives $M^2 \propto -i\omega + \frac12 D_x k^2_x+\frac12 D_\perp k^2_A$. Then the full Green's function should be
\begin{equation}
G^R_{nn}(\omega,k)
	= \frac{\chi \omega}{\omega + i D_\perp k_\perp^2 + i  k_x^2  \left\{ - \frac{3 \chi T c^{'2}}{32 \pi } \log \frac{-i\omega + \frac12 D_x k_x^2+\frac 12 D_\perp k_y^2}{\Lambda^2}\right\}^{2/3}}\ .
\end{equation}
Returning to the lab frame $\omega\to \omega + c k_x$, we see that in addition to the original pole $\omega = ck_x - iD_{ij} k^i k^j$, response functions feature `two-diffuson' branch cuts with branch points located at $\omega_{\hbox{\scriptsize 2-diff}} = ck_x - \frac{i}{2}D_{ij} k^i k^j$. Higher loop contributions with cuts involving $n$ excitations will similarly lead to additional branch points $\omega_{\hbox{\scriptsize $n$-diff}} = ck_x - \frac{i}{n}D_{ij} k^i k^j$. This analytic structure was discussed in the case of simple diffusion (i.e.~with $c=0$) in \cite{Chen-Lin:2018kfl}.

Finally, from the scalings obtained below Eq.~\eqref{conss1}, interactions are irrelevant for spatial dimensions $d>2$, where the coupling $c'$ scales as $1/\Lambda^{\frac{d-2}{2}}$. In this case resumming the geometric series of diagrams from Fig. \ref{loop} gives
\begin{equation}
G^R_{nn}(\omega,k)
	= \frac{\chi \omega}{\omega + i D_{ij}k^ik^j + i \Sigma(\omega,k)}\,  ,\qquad
\hbox{with} \quad
\Sigma(\omega,k)
	= \frac{c'^2 T k_x^2}{16\chi \sqrt{(8\pi D_x)(8\pi D_\perp)^{d-1}}} \frac{\left(-i\omega + \frac{1}{2} D_{ij} k^i k^j\right)^{\frac{d}{2}-1}}{\Gamma \left(\frac{d}{2}\right) \sin \left(\frac{d\pi}{2}\right)}  \ .
\end{equation}
For $d$ even, the expression should be expanded around $d$ like above, dropping the $1/\sin\left(\frac{d\pi}{2}\right)$ divergence, and one finds an additional logarithmic enhancement (this logarithmic enhancement can also be obtained directly from \eqref{eq_dimreg} by using a momentum cutoff instead of dimensional regularization). Specializing to $d=3$, relevant to the chiral magnetic effect, one finds a long-time tail correction to transport along the magnetic field (dropping numerical factors):
\begin{equation}
\sigma_{xx}(\omega) \sim \sigma_{xx}(0) + \frac{c'^2 T}{\sqrt{D_x D_\perp^2}} \omega^{1/2} + \cdots
	\, .
\end{equation}
%


\end{document}